\documentclass[12pt]{article}
\usepackage[utf8]{inputenc}
\usepackage[left=2cm, right=2cm, top=2cm, bottom=2cm]{geometry}

\title{The numerical simulations for forecast of the epidemic COVID-19 models in US, Brazil, South Korea, India, Russia and Italy}
\author{Chun-sheng Chen$^{*}$, Bo-Cyuan Lin$^{*}$, Yen-Jia Chen$^{\dagger}$,\and  Yi-Cheng Hung$^{\dagger}$, Han-Chun Wang$^{*}$ and Jann-Long Chern$^{\dagger}$$^{\ddagger}$ \and \scriptsize{$^{*}$ Department of Mathematics, National Central University, Taoyuan, Taiwan.} \and \scriptsize{$^{\dagger}$ Department of Mathematics, National Taiwan Normal University, Taipei, Taiwan.} \and \scriptsize{$^{\ddagger}$ Corresponding author: 
  The authors are supported in part by Ministry of Science and Technology (MOST), Taiwan,} \and \scriptsize{No. MOST 107-2115-M008-005-MY3, and 2020 NCTS USRP, Taiwan. chern@math.ntnu.edu.tw}}
\date{}

\usepackage{mathtools}
\usepackage[ruled,vlined]{algorithm2e}
\usepackage{algorithmic}
\usepackage{amsmath,amsthm,amssymb}
\DeclareMathOperator*{\argmin}{arg\,min}
\usepackage[caption=false]{subfig}
\theoremstyle{definition}
\newtheorem*{key}{Keywords}
\newtheorem{theorem}{Theorem}[section]

\begin{document}

\maketitle

\begin{abstract}
The main purpose of this paper is to present the numerical simulations for forecast of the epidemic COVID-19 models in US, Brazil, South Korea, India, Russia and Italy. By using the SQIARD and SIARD  models of COVID-19 with quarantine and asymptomatic infected, we respectively apply the infection data and adjust the related parameters in numerical simulations to generate the forecast and eﬀect of prediction of the epidemic situation for each country. At the same time, we also use the US infection data to compare SQIARD with SIARD and show the effect of its prediction.
\end{abstract}

\begin{key}
\ \\  
COVID-19, quarantine, asymptomatic infection, SIARD Model, SQIARD Model, data forecast
\end{key}


\section{Introduction}
As the data in WHO report (WHO Coronavirus (COVID-19) Dashboard, as of 6:46pm CEST, 20 May 2021), globally there have been 164,523,894 confirmed cases of COVID-19, including 3,412,032 deaths. COVID-19 become the most serious infectious disease and has had a dramatic impact on the human health and also damage the worldwide economic and developments now. As the real status which COVID-19 patients have high percentage with mild symptoms or no symptoms. The actual symptomatic cases are much lower than the pre-estimating. In \cite{intro1}, the author doing the estimate for 11 European countries, lot of peoples had been infected but the patients has been detected are much less than actual infections. Two major reasons to caused this problem. One is limited testing capacity and another is high percentage of mild symptoms or no symptoms. Lavezzo, et al \cite{intro2} stated that, at Vò, Italy, the asymptomatic cases were 43.2\% of the total. In \cite{intro3}, based on government estimations, the article assumed that the numbers of asymptomatic infected patients is nine times higher than the numbers of symptoms patients. Gudbjartsson, et al \cite{intro4} also pointed out that the detected data of SARS-CoV-2 which done by screening group of Iceland and also showed 43\% of the participants with asymptomatic. In \cite{intro5},  the percentage of asymptomatic patients increased very fast from 16.1\%(35 asymptomatic infections/218 conﬁrmed cases) to 50.6\%(314/621) within a week. As this fact, the proportion of asymptomatic infections for COVID-19 has dramatic changed. As the WHO's report, it pointed out that the asymptomatic patients are not non-infectious. Therefore, the proportion of asymptomatic infections is very critical to the impact of epidemic research. 

This paper will use mathematical models and data analysis methods to analyze the proportion of asymptomatic infections in various countries and predict the future trend of the epidemic. In Table 1 of section 4,we list the symptomatic infections with $\alpha$ in six countries. $1-\alpha$ is the proportion of asymptomatic people. In \cite{intro2}, the authors also stated that, at V‘o, Italy, the proportion of asymptomatic cases are 43.2\% which was consistent with the 45\% of our numerical simulation in the Table 1 for Italy. This also shows that our model can be applied to ﬁnd out the proportion of asymptomatic people in each country

In the study of the COVID-19 epidemic, one of important topics is to estimate the basic reproduction number $R_{0}$. In \cite{intro7} and \cite{intro8}, the relation between the model locally asymptotically stability and $R_{0}$ had been found. In \cite{intro9}, the relation between the globally asymptotically stability of the model and $R_{0}$ by the authors. In the best results of \cite{intro10}, Chen, et al revised the respective model into discrete time difference equations to find the relation between $R_0$ and model parameters. The relation between the reproduction numbers and sub-threshold en-demic  equilibria  for compartmental models of disease transmission we refer the nice and interesting paper \cite{intro12} and et al. In our article, we will revise the differential systems into discrete time diﬀerence equations to train the model parameters for the epidemic prediction respectively.

In the previous SARS epidemic, many researchers have discussed the modifications of the SARS epidemic in terms of ``quarantine" and ``asymptomatic", e.g., please refer to \cite{intro7}-\cite{intro8} and the related references. However, because SARS and COVID-19 have different epidemic patterns, the disease patterns of asymptomatic infections are also different. Therefore, we establish two mathematical models, SQIARD and SIARD model, to simulate the COVID-19 epidemic. Obviously the SIARD model is only a simpliﬁed form of the SQIARD model. The difference between this two models are one include the parameters ``number of people to be screened for the epidemic" and another one exclude it.  It is very hard to get the data for this parameter but fortunately US government has complete data on the number of people in quarantine. This data can be used to generate the model parameters. In order to get the model parameters, we remove the parameter $Q(t)$ first and use SIARD model to conduct the prediction data of US, South Korea, Brazil, India, Russia and Italy. The prediction of the SIARD model will be created first and then use US's data which contains $Q(t)$ to show the effect of prediction of the SQIARD model.

The organization of this article are as follows:  In Section 2, we utilize SQIARD to investigate the transmission model of COVID-19 with quarantine, infected and asymptomatic infected, and introducing the corresponding numerical simulation algorithm. In Section 3, we first create SIARD model and use FIR algorithm to generate the respective parameters of model.  Finally, in Section 4, we present the forecast data of US, South Korea, Brazil, India, Russia and Italy respectively. Meanwhile, we also review the effect of prediction of the epidemic situation in each country, and use US’s data to compare SQIARD with SIARD and show the effects of predictions

\section{SQIARD Model}
In this section, we will apply SQIARD to investigate the transmission model of COVID-19 with quarantine, infected and asymptomatic infected, and introducing the corresponding numerical simulation algorithm. 
\subsection{The Derivation of Model and Basic Reproduction Number}
In SQIARD model, the normal people can be infected by infected patients and asymptomatic patients. Those peoples who with negative quarantine will be screened. It create three types after screening: symptomatic, asymptomatic, and negative. In the end, the infected person will be gradually moved to the class of recovery or death.
The model has the following assumptions:
\begin{enumerate}
    \item $Q(t)$ is the number of quarantine(daily screening) people per day. Those with positive quarantine results are further divided into ``symptomatic patients"($I(t)$) or ``asymptomatic patients"($A(t)$). Those with negative quarantine results will return to $S(t)$ at a rate of $\mu_{3}$.
    \item $A(t)$ is claimed to be had less infectivity, $0 < \delta <1$, where $\delta$ is the proportion in infectiousness of asymptomatic infectives.
    \item Total Population in our model are viewed as the same. 
    \item We simply the total population as a fixed value without new born and non-epidemic death.
    \item There is no infectiousness during the daily quarantine (screening) process.
\end{enumerate}

The variables are given as follows: [$S$: susceptible population; $Q$: quarantine population; $I$: infective population; $A$: asymptomatic infective population; $R$: recovered population; $D$: deaths].

\begin{figure}[htp]
    \centering
    \includegraphics[width=8 cm]{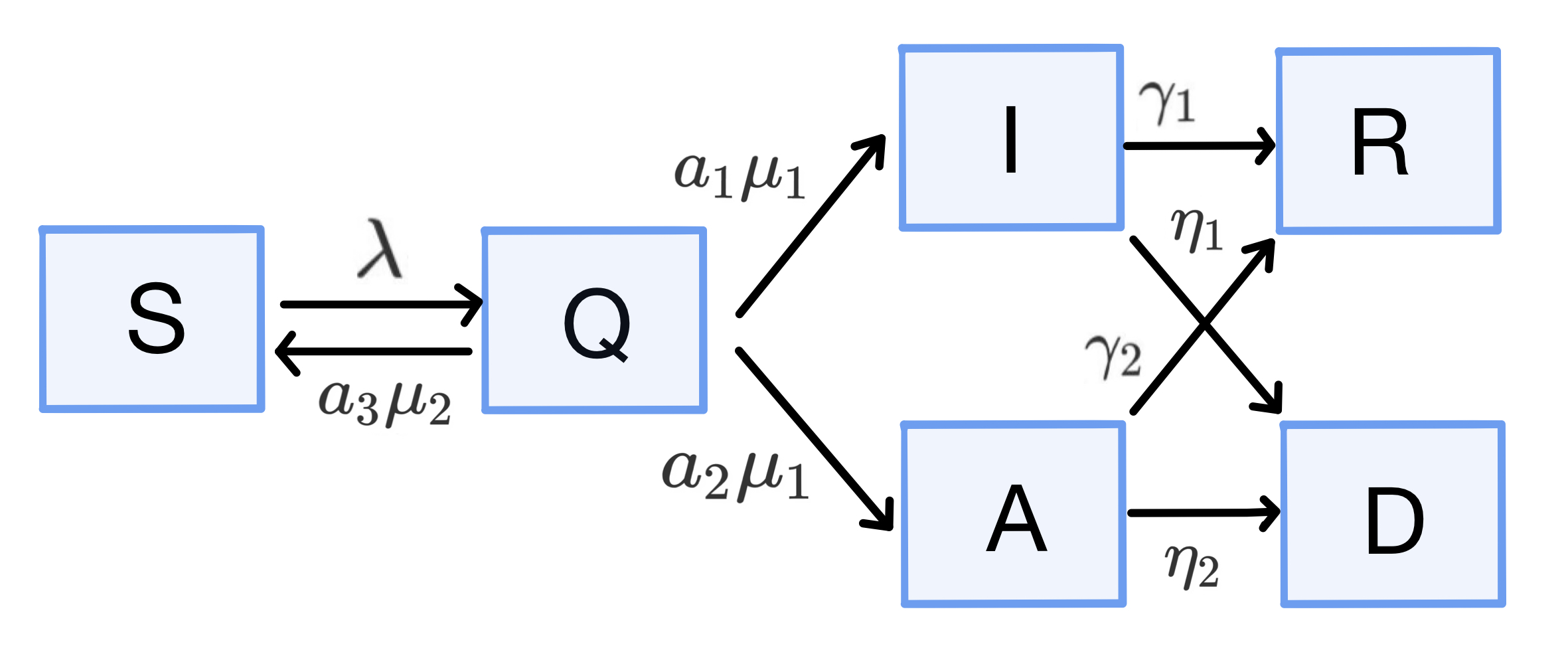} 
    \caption{The flow chart of SQIARD}
\end{figure}
\noindent
The model parameters are given as follows: [$\beta$: the progression rate of susceptible to quarantine classes; $\delta$: the proportion in infectiousness of asymptomatic infectives, where $0<\delta<1$; $\mu_1$: the progression rate of susceptible to infective classes; $\mu_2$: the progression rate of quarantine to susceptible classes; $a_i$: with $0<a_i<1$, the proportion of susceptible class $Q$ progressing to positive class $I$, $A$, or negative class which will back to the susceptible $S$, and $a_1+a_2+a_3=1$; $\gamma_1$ and $\gamma_2$: the recovered rates of infective classes $I$ and $A$; $\eta_1$ and $\eta_2$: are the disease death rates of infective classes $I$ and $A$].\\
From above, $S(t)$ , $Q(t)$ , $I(t)$ , $A(t)$ , $R(t)$ and $D(t)$ are conducted by the following differential equations:

\begin{equation} \label{eq:SQIARD}
\begin{cases}
S^{'} = -\lambda S +a_3\mu_2 Q\\
Q^{'} = \lambda S -\left(a_1\mu_1+a_2\mu_1+a_3\mu_2 \right)Q\\
I^{'} = a_1 \mu_1 Q -(\gamma_1 +\eta_1)I\\
A^{'} = a_2\mu_1 Q - (\gamma_2 +\eta_2)A\\
R^{'} =\gamma_1 I+\gamma_2 A\\
D^{'} = \eta_1 I +\eta_2 A,\\
\end{cases}
\end{equation}
where $\lambda =\dfrac{\beta(I+\delta A)}{(S+I+A)}$ , $S(0)=S_0>0$ , $Q(0)=0$ , $I(0)=I_0>0$ , $A(0)=A_0>0$ , $R(0)=0$ and $D(0)=0$. 
We note that $S(t)+Q(t)+I(t)+A(t)+R(t)+D(t)=N=S_0+I_0+A_0$ where $N$ is the total population. Then the corresponding basic reproduction number of (\ref{eq:SQIARD}) is:
\begin{equation}
R_Q=\dfrac{\mu_1}{a_1\mu_1+a_2\mu_1+a_3\mu_2}[\dfrac{a_1}{\gamma_1+\eta_1}+\dfrac{a_2\delta}{\gamma_2+\eta_2}]\beta.    
\end{equation}
\\
Due to daily updating of the COVID-19, we revise (\ref{eq:SQIARD}) into discrete form (\ref{SRD}) and (\ref{IAQ}):
\begin{equation} \label{SRD}
    \begin{array}{l}
    S(t+1)-S(t)=-\dfrac{\beta S(I+\delta A)}{S+I+A}+a_3\mu_2Q(t)\\
    \ \\
    R(t+1)-R(t) = \gamma_1 I+\gamma_2 A\\
    \ \\
    D(t+1)-D(t) = \eta_1 I + \eta_2 A.\\
    \end{array}
\end{equation}
When the disease first spreads, the number of people infected is much smaller than the total population. The total population can be assumed to the same as the number of suspected infections and let $q=(a_{1}\mu_{1}+a_{2}\mu{1}+a_{3} \mu_{2})$ further simplify as follows :
\begin{equation} \label{IAQ}
    \begin{array}{l}
    I(t+1)-I(t)=a_1\mu_1Q(t) -\gamma_1 I -\eta_1 I\\
    \ \\
    A(t+1)-A(t)=a_2\mu_1Q(t) -\gamma_2 A -\eta_2 A\\
    \ \\
    Q(t+1)-Q(t)=\beta(I+\delta A)-qQ(t).\\ 
    \ \\
    \end{array}
\end{equation}

\subsection{Tracking Time-Depend SQIARD Model Algorithms}
The Covid-19 data from the WHO is discrete time-series. For conducting the prediction, we make the following assumptions which are different from the original model.
\begin{enumerate}
    \item Since the data of Covid-19 contains $\{S(t), Q(t), \mathcal{I}(t),$ $R(t), D(t), \ 0 \leq t \leq T \}$ without asymptotic infectious, we consider $A(t) = \alpha*\mathcal{I}(t),$ \ $I(t) = (1-\alpha)*\mathcal{I}(t)$, where $\alpha = 0.6$ (referred from \cite{intro13}, \cite{intro14}), $\delta = 0.5$ (referred from \cite{intro7}).
    \item Rewrite the differential equations (section 2.1 (1)) as discrete form (5) similar with the consideration in paper \cite{intro10}. Then, in (5), what data we substitute into the block I, A, R, D is the data obtained by the WHO (i.e $S$ contains the other infectious, asymptotic infectious, recovery and death which is not confirmed to collect to the data).
    \item The component $\frac{\beta(t)S(t) (I(t) + \delta A(t))}{S(t) + I(t) + A(t)}$ of equation $S(t+1)$, is conducted by $I(b), A(b), b < t$, it means that the susceptible population has already contacted with the infectious population(time $t$) before time $t$.
\end{enumerate}
Base on above assumptions, we implement the predicting algorithm and provide our predicting result in the section 4.
\begin{equation}
\begin{cases}
    S(t+1)&= S(t) - \displaystyle \frac{\beta(t)S(t) (I(t) + \delta A(t))}{S(t) + I(t) + A(t)} + a_{3}(t) \mu_{2}Q(t)\\
    Q(t+1)&= Q(t) + \displaystyle \frac{\beta(t)S(t) (I(t) + \delta A(t))}{S(t) + I(t) + A(t)} \\
    &- (a_{1}(t) \mu_{1} + a_{2}(t) \mu_{1}+a_{3}(t) \mu_{2}) Q(t)\\
    I(t+1)&= I(t) + a_{1}(t) \mu_{1}Q(t) - \gamma_{1}(t)I(t) - \eta_{1}(t)I(t)\\
    A(t+1)&= A(t) + a_{2}(t) \mu_{1}Q(t) - \gamma_{2}(t)A(t) - \eta_{2}(t)A(t)\\
    R(t+1)&= R(t) + \gamma_{1}(t)I(t) + \gamma_{2}(t)A(t)\\
    D(t+1)&= D(t) + \eta_{1}(t)I(t) + \eta_{2}(t)A(t)
\end{cases}
\end{equation}

 If the asymptomatic infected people die in a very low probability, then we can consider $\eta_{2} = 0$. Then, in this case, since the relation $a_{1} + a_{2} + a_{3} = 1$, the $\beta (t), \gamma_{1} (t), \gamma_{2} (t), \eta_{1} (t), a_{3} (t)$ and the time-depend basic reproduction number of SQIARD  can be evolved as following (\ref{SQIRDmodels}) and (\ref{rq}):

\begin{equation} \label{SQIRDmodels}
\begin{aligned}
    \beta(t) & = \frac{(S(t) - S(t+1) + a_{3}\mu_{2}Q(t))}{S(t)(I(t) + \delta A(t))} (S(t) + I(t) + A(t))\\
    \gamma_{1}(t) & = \frac{I(t) - I(t+1) + a_{1} \mu_{1} Q(t)}{I(t)}\\
    \gamma_{2}(t) & = \frac{R(t+1) - R(t) + I(t+1) - I(t) + a_{1} \mu_{1} Q(t)}{A(t)}\\
    \eta_{1}(t) & = \frac{D(t+1)-D(t)}{I(t)}\\
    a_{3}(t) & = \frac{S(t+1)-S(t)+Q(t+1)-Q(t)+\mu_{1}Q(t)}{\mu_{1}Q(t)}
\end{aligned}
\end{equation}

\begin{small}
\begin{equation} \label{rq}
R_{Q}(t) = \frac{\beta(t) \mu_{1}}{(a_{1}(t)\mu_{1}+a_{2}(t)\mu_{1}+a_{3}(t)\mu_{2})}( \frac{a_{1}(t)}{\gamma_{1}(t) + \eta_{1}(t)}+ \frac{\delta a_{2}(t) }{\gamma_{2}(t)}).
\end{equation}
\end{small}
\noindent
We use Finite Impulse Response filters (FIR) (8), to predict $\hat{\beta}(t),\hat{\gamma_{1}}(t),\hat{\gamma_{2}}(t),\hat{\eta_{1}}(t), \hat{a_{3}}(t)$. We also note that $\hat{a}_{1}(t), \hat{a}_{2}(t)$ can be obtained from $\hat{a}_{3}(t)$.
\begin{equation} \label{fir}
\begin{aligned}
    \hat{\beta} (t) & = a_{0} + a_{1} \beta (t-1) + \cdots + a_{J_{1}} \beta (t-J_{1}) = \sum_{j=1}^{J_{1}} a_{j} \beta(t-j) + a_{0}\\
    \hat{\gamma_{1}} (t) & = b_{0} + b_{1} \gamma_{1} (t-1) + \cdots + b_{J_{2}} \gamma_{1} (t-J_{2}) = \sum_{j=1}^{J_{2}} b_{j}\gamma_{1}(t-j) + b_{0}\\
    \hat{\gamma_{2}} (t) & = c_{0} + c_{1} \gamma_{2} (t-1) + \cdots + c_{J_{3}} \gamma_{2} (t-J_{3}) = \sum_{j=1}^{J_{3}} c_{j}\gamma_{2}(t-j) + c_{0}\\
    \hat{\eta_{1}} (t) & = d_{0} + d_{1} \eta_{1} (t-1) + \cdots + d_{J_{4}} \eta_{1} (t-J_{4}) = \sum_{j=1}^{J_{4}} d_{k}\eta_{1}(t-j) + d_{0}\\
    \hat{a_{3}} (t) & = e_{0} + e_{1} a_{3} (t-1) + \cdots + e_{J_{5}} a_{3} (t-J_{5}) = \sum_{j=1}^{J_{5}} e_{k}a_{3}(t-j) + e_{0},
\end{aligned}
\end{equation}
where $a_{j_{1}}, \ j_{1} = 0, 1, ..., J_{1}; \ b_{j_{2}}, \ j_{2} = 0, 1, ..., J_{2};$ $\ c_{j_{3}}, \ j_{3} = 0, 1, ..., J_{3}; \ d_{j_{4}}, \ j_{4} = 0, 1, ..., J_{4}; \ e_{j_{5}}, \ j_{5} = 0, 1, ..., J_{5}$ are the coefficients (weight) of the five given FIR filters as above.
\noindent
We will adopt the following Ridge Regularization method (\ref{ridge}) which is often used in the machine learning for each FIR models, and use Theorem \ref{thm:nge} implemented by Algorithm \ref{alg:nge} to optimize the respective weights.
\begin{equation}\label{ridge}
\begin{aligned}
    \min_{ \{ \mathbf{a} \} } F_{\beta}(\mathbf{a}) & =  \min_{ \{ \mathbf{a} \} }[\sum_{t = J_{1}}^{T-2}(\beta(t) - \hat{\beta}(t))^{2} + m_{1}\sum_{i=0}^{J_{1}} a_{i}^{2}]\\
    \min_{ \{ \mathbf{b} \} } F_{\gamma_{1}}(\mathbf{b}) & = \min_{ \{ \mathbf{b} \} }[\sum_{t = J_{2}}^{T-2}(\gamma_{1}(t) - \hat{  \gamma_{1}}(t))^{2} + m_{2}\sum_{i=0}^{J_{2}} b_{i}^{2}]\\
    \min_{ \{ \mathbf{c} \} } F_{\gamma_{2}}(\mathbf{c}) & = \min_{ \{ \mathbf{c} \} }[\sum_{t = J_{3}}^{T-2}(\gamma_{2}(t) - \hat{\gamma_{2}}(t))^{2} + m_{3}\sum_{i=0}^{J_{3}} c_{i}^{2}]\\
    \min_{ \{ \mathbf{d} \} } F_{\eta_{1}}(\mathbf{d}) & = \min_{ \{ \mathbf{d} \} }[\sum_{t = J_{4}}^{T-2}(\eta_{1}(t) - \hat{\eta_{1}}(t))^{2} + m_{4}\sum_{i=0}^{J_{4}} d_{i}^{2}]\\
    \min_{ \{ \mathbf{e} \} } F_{a_{3}}(\mathbf{e}) & = \min_{ \{ \mathbf{e} \} }[\sum_{t = J_{5}}^{T-2}(a_{3}(t) - \hat{a_{3}}(t))^{2} + m_{5}\sum_{i=0}^{J_{5}} e_{i}^{2}], 
\end{aligned}
\end{equation}
where $\mathbf{a} = (a_{0}, a_{1}, ..., a_{J_{1}}), \ \mathbf{b} = (b_{0}, b_{1}, ..., b_{J_{2}}), \ \mathbf{c} = (c_{0}, c_{1}, ..., c_{J_{3}}), \ \mathbf{d} = (d_{0}, d_{1}, ..., d_{J_{4}}), \ \mathbf{e} = (e_{0}, e_{1}, ..., e_{J_{4}}).$

\subsubsection{Normal Gradient Equation}
Before processing our numerical algorithm, we need the following theorem. 
\begin{theorem}
\label{thm:nge}
(Normal Gradient Equation)\\
Let $f (0), f(1), ..., f(T-2)$ be the training data with $T-1$ points, and the FIR filter $\hat{f} (t) = x_{0} + x_{1} f (t-1) + \cdots + x_{J} f(t-J)$ be the prediction of $t-$th point, $t = J, ..., T-2$ with cost function $\textstyle F(x_{0}, x_{1}, ..., x_{J};m) = \sum_{t=J}^{T-2}(f(t) - \hat{f}(t))^{2} + m \sum_{i=0}^{J} x_{i}^{2}$, where $m$ is the regression parameter. If $m \notin \sigma(-A)$ then $\mathbf{x_0}:=(x_{0}, x_{1}, ..., x_{J})^{T} = (m I + A)^{-1} b$ satisfies $F(\mathbf{x_0};m) = \displaystyle\min_{\mathbf{x\in R^{J+1}}} F(\mathbf{x};m)$ where $A_{(J+1)\times (J+1)}, \ mI_{(J+1) \times (J+1)}$ and $b_{(J+1) \times 1}$ are defined as follows:
\begin{tiny}
$$
A = \begin{bmatrix}
(T-J-1)& \displaystyle\sum^{T-2}_{t=J}f(t-1) & \cdots&\displaystyle\sum^{T-2}_{t=J}f(t-J)\\
\displaystyle\sum^{T-2}_{t=J}f(t-1)&\displaystyle\sum^{T-2}_{t=J}f(t-1)^{2} &\cdots&\displaystyle\sum^{T-2}_{t=J}f(t-1)f(t-J)\\
\vdots&\vdots&\ddots&\vdots\\
\displaystyle\sum^{T-2}_{t=J}f(t-J)&\displaystyle\sum^{T-2}_{t=J}f(t-J)f(t-1)&\cdots&\displaystyle\sum^{T-2}_{t=J}f(t-J)^{2}
\end{bmatrix},
$$
\end{tiny}
\begin{tiny}
$$
m I = 
\begin{bmatrix}
    m & 0 & 0 & \cdots & 0\\
0 & m & 0 & \cdots & 0\\
0 & 0 & m & \cdots & 0\\
\vdots&\vdots&\vdots&\ddots&\vdots\\
0 & 0 &\cdots&\cdots\cdots&\ m\\ 
\end{bmatrix}, \ 
b = 
\begin{bmatrix}
\sum^{T-2}_{t=J}f(t)\\
\sum^{T-2}_{t=J}f(t)f(t-1)\\
\sum^{T-2}_{t=J}f(t)f(t-2)\\
\vdots\\
\sum^{T-2}_{t=J}f(t)f(t-J)
\end{bmatrix}.
$$
\end{tiny}

\begin{proof} It is easily to see that if $F(\mathbf{x_0};m) = \displaystyle\min_{\mathbf{x\in R^{J+1}}} F(\mathbf{x};m)$ then $\textstyle \frac{\partial}{\partial x_{j}} F(\mathbf{x_0};m) = 0\ \forall j=0\cdots J$.
Hence at the minimal point $\mathbf{x_0}$ we have:

\begin{scriptsize}
\begin{equation*}
\begin{aligned}
\frac{\partial}{\partial x_{0}} F(\mathbf{x}_{0};m)
&= \frac{\partial}{\partial x_{0}} [\sum_{t=J}^{T-2}(f(t) - \hat{f}(t))^{2} + m \sum_{i=0}^{J} x_{i}^{2}] \\
&= -2 \sum_{t=J}^{T-2} ](f(t) - (x_{0} + x_{1} f(t-1)\\
&+ \cdots + x_{J} f (t-J))) \cdot 1] \\
&+ 2 m x_{0},\\
\frac{\partial}{\partial x_{j}} F(\mathbf{x}_{0};m)&=\frac{\partial}{\partial x_{j}} [\sum_{t=J}^{T-2}(f(t) - \hat{f}(t))^{2} + m \sum_{i=0}^{J} x_{i}^{2}] \\
&= -2 \sum_{t=J}^{T-2}[(f(t) - (x_{0} + x_{1} f(t-1)] + \cdots + x_{J} f (t-J)))\\
& \cdot f(t-j)+ 2 m x_{j}, \ j=1,2,..,J.
\end{aligned}
\end{equation*}
\end{scriptsize}

Thus we easily obtain the following results:
\begin{scriptsize}

\begin{align*}
    \displaystyle\sum_{t=J}^{T-2} f (t) &= \sum_{t=J}^{T-2} (x_{0} + x_{1} f (t-1) + \cdots + x_{J} f(t-J)) + m x_{0}, \\
    \displaystyle\sum_{t=J}^{T-2} f (t) &= \sum_{t=J}^{T-2} (x_{0} + x_{1} f (t-1) + \cdots + x_{J} f(t-J)) \cdot f (t-j)\\
    &+ m x_{j}, \ j \neq 0.    
\end{align*}
    
\begin{tiny}
$$
\Rightarrow
\begin{bmatrix}
(T-J-1) + m & \displaystyle\sum^{T-2}_{t=J}f(t-1) & \cdots&\displaystyle\sum^{T-2}_{t=J}f(t-J)\\
\displaystyle\sum^{T-2}_{t=J}f(t-1)&\displaystyle\sum^{T-2}_{t=J}f(t-1)^{2} + m &\cdots&\displaystyle\sum^{T-2}_{t=J}f(t-1)f(t-J)\\
\vdots&\vdots&\ddots&\vdots\\
\displaystyle\sum^{T-2}_{t=J}f(t-J)&\displaystyle\sum^{T-2}_{t=J}f(t-J)f(t-1)&\cdots&\displaystyle\sum^{T-2}_{t=J}f(t-J)^{2} + m
\end{bmatrix} 
\begin{bmatrix}
    x_{0}\\
    x_{1}\\
    \vdots\\
    x_{J}\\
\end{bmatrix}
$$
\end{tiny}
$$
=\begin{bmatrix}
\sum^{T-2}_{t=J}f(t)\\
\sum^{T-2}_{t=J}f(t)f(t-1)\\
\vdots\\
\sum^{T-2}_{t=J}f(t)f(t-J)
\end{bmatrix}
$$
$$
\Rightarrow (x_{0}, x_{1}, ..., x_{J})^{T} = (m I + A)^{-1} b\hbox{ if } m \notin \sigma(-A).
$$
\end{scriptsize}
This completes the proof.
\qed \end{proof}
\end{theorem}
{\bf Remark 2.1} We note that, for giving a regression parameter $m>0$, the cost function $F$ is positive of degree $2$ with respect to each $x_j$ and   $F(\mathbf{x};m)\to\infty$ as $x_j\to\pm\infty,~ \mathbf{x}=(x_0, x_1,\cdots, x_J).$ It follows that $F(\mathbf{x_0};m)= \displaystyle\min_{\mathbf{x\in R^{J+1}}} F(\mathbf{x};m)$ if and only if $\frac{\partial}{\partial x_{j}} F(\mathbf{x_0};m) = 0\ \forall j=0\cdots J$.\\

From Theorem \ref{thm:nge} we have the following numerical {\bf Algorithm 2.1.}
\begin{algorithm}
\caption{Normal Gradient Equation}
\label{alg:nge}
\begin{algorithmic}
\STATE{\textbf{Input:} Training Data $\{f (t),0\leq t\leq T-2\}$; Regularization Parameter $m$; Order of FIR $J$.}
\STATE{Compute $A, b$ of $(m I +A)\mathbf{x}=b$ with cost function and ridge regularization as theorem above.}
\STATE{Choose $m \notin \sigma(-A)$ if not change $m$.}
\STATE{Solve $\mathbf{x}=(m I +A)^{-1}b$.}

\RETURN Weights of FIR Model: $\mathbf{x}= \{x_i,0\leq i \leq J\}$
\end{algorithmic}
\end{algorithm}
\subsubsection{Option for the Orders of FIR filters and Reference Effective Interval}
In order to find the appropriate orders for training each FIR models ($\hat{\beta}(t), \hat{\gamma_{1}}(t),$
$ \hat{\gamma_{2}}(t), \hat{\eta_{1}}(t), \hat{a_{3}}(t)$), we divide the parameters data $\{ \beta(t), \gamma_{1}(t), \gamma_{2}(t), \eta_{1}(t), a_{3}(t),$ $\ 0 \leq t \leq T-2 \}$ into respective two parts, the training data set (size: $\ell_{T}$) and the validation set (size: $\ell_{V}$) where $\ell_{V} = T-1 - \ell_{T}$.
\begin{algorithm}
\caption{Order Searcher}
\label{alg:ordersearcher}
\begin{algorithmic}
\STATE{\textbf{Input:} Data: $\{ f(t), 0 \leq t \leq  T-2\}$; Training size: $\ell_{T}$; Lower Bound of Order: $L_{J}$; Upper Bound of Order: $U_{J}$; Regularization Parameter: $m$.}
\STATE{Divide the data into Training Set: $Data_{T} := \{ f(t), 0 \leq t \leq \ell_{T}-1 \}$, and Validation Set: $Data_{V} := \{ f(t), \ell_{T} \leq t \leq T-2 \}$.}
\STATE{Compute Validation Length: $\ell_{V}$.}
\FOR{$J \gets L_{J}$ \textbf{to} $U_{J}$}
\WHILE{$\ell_{T} \leq t \leq T-2$}
\STATE{Train model with (\ref{ridge}), $J$, $m$ and $Data_{T}$ by Algorithm \ref{alg:nge}}
\STATE{Estimate $\hat{f_{J}}(t)$, then Append to Predicting set $Pred_{V}(J):= \{ \hat{f_{J}}(t), \ell_{T} \leq t \leq T-2 \}$ of Validation.}
\ENDWHILE
\STATE{Calculate $\displaystyle err(J) := \sum_{t} |Pred_{V}(J) - Data_{V}| = \sum_{t = \ell_{T}}^{T-2} |\hat{f_{J}}(t) - f(t)|$, then Append to Error Set $Error:= \{ err(J), L_{J} \leq J \leq U_{J} \}$.}
\ENDFOR
\STATE{$J_{fit} = \displaystyle\argmin_{ \{ J \} }[Error]$.}
\RETURN Orders of FIR: $J_{fit}$.
\end{algorithmic}
\end{algorithm}

By using the training set for fitting the model, we take the prediction with the same length as the validation set for different orders on certainly suitable range $([L_{J},U_{J}] \subseteq \mathbb{N})$. After this step, we can find the argument of the minimum for the sums of errors w.r.t each orders $J$ as follows

$$
\displaystyle \argmin_{\{ J \}}[\sum_{t = \ell_{T}}^{T-2} |\hat{f_{J}}(t) - f(t)|], \  L_{J} \leq J \leq U_{J}.
$$
In Figure 2, we give a illustration of the result for the forecast based on the three algorithms. As implementing the Algorithm 2 to find the fit order, we also compute the reference effective interval 5\%, 10\%, 20\% which are intervals of days satisfies $\frac{(predicting) - (real data)}{real data} < 5\%, 10\%, 20\%$ in the validation set.

\begin{figure}[htp]
    \centering
    \includegraphics[width=10cm]{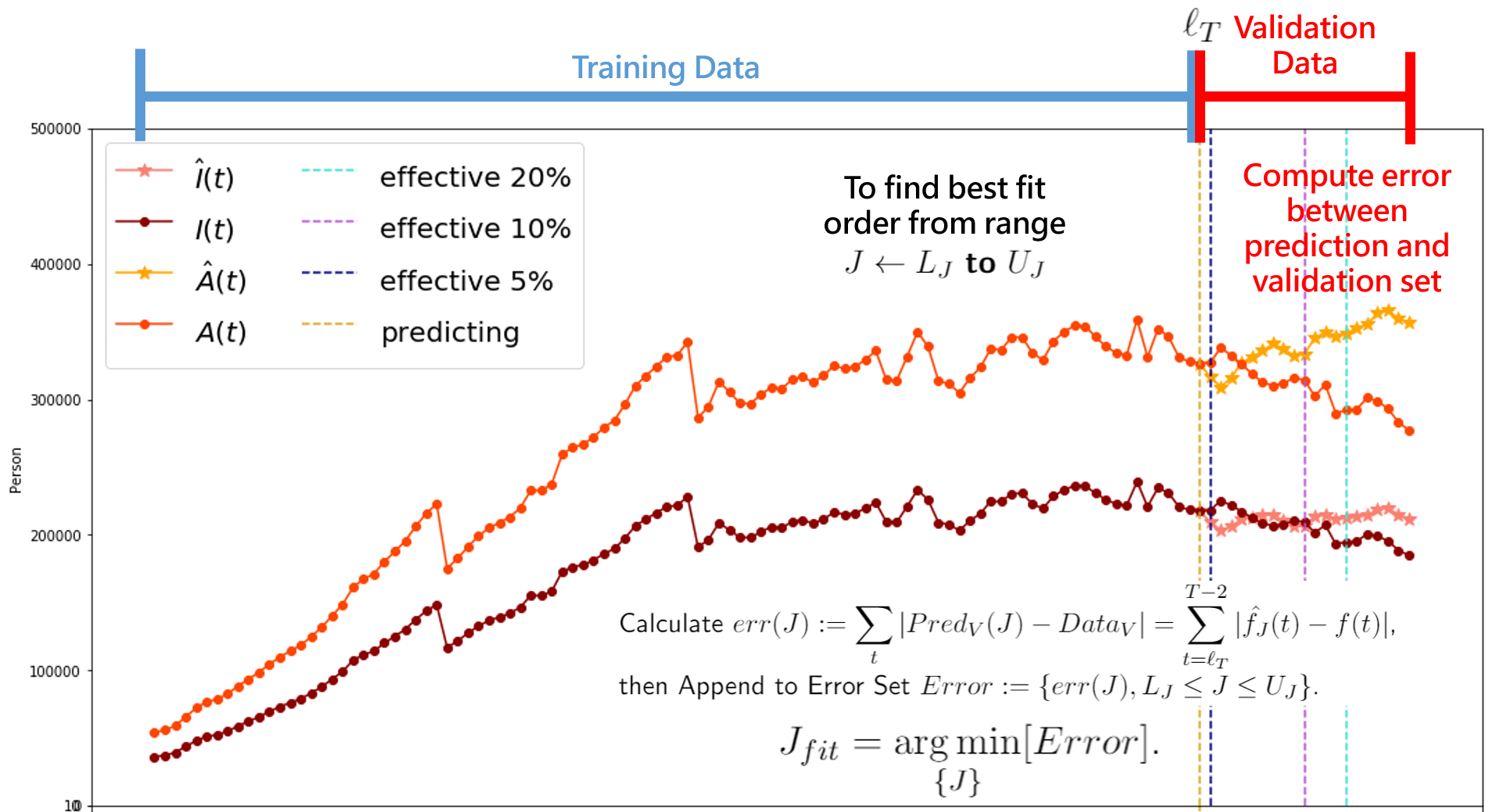}
    \caption{Reference Effective Interval with validation set}
    \label{fig:interval}
\end{figure}
Before conducting the tracking Algorithm \ref{alg:trackingSQIARD}, we already obtain the respective order of $\hat{\beta}$, $\hat{\gamma_{1}}$, $\hat{\gamma_{2}}$, $\hat{\eta_{1}}$, $\hat{a_{3}}$ by Algorithm \ref{alg:ordersearcher} and the reference effective intervals for the forecast.
To increase the variety, append the prediction $\hat{\beta} (t),\hat{\gamma_{1}} (t),\hat{\gamma_{2}} (t),\hat{\eta_{1}} (t), \hat{a_{3}}(t), \geq T-1$ to the training set as following.

Let $f(t), \ 0 \leq t \leq  T-2$ be the training set, $\hat{f}(t), \ t \geq T-1$ be the prediction depend on the trained model and $P$ be the stopping criteria of the forecast. Then the Appended Training Data defined as: $\mathcal{F} := \{ \mathcal{F}(t) | (\mathcal{F}(t) = f(t), \ if \ 0 \leq t \leq T-2) \ or \ (\mathcal{F}(t) = \hat{f}(t), \ if \ T-1 \leq t \leq T+P)  \}$, e.g., the Appended training data set of $\beta,~ \mathcal{B} := \{ \mathcal{B}(t) | (\mathcal{B}(t) = \beta (t), \ if \ 0 \leq t \leq T-2) \ or \ (\mathcal{B}(t) = \hat{\beta}(t), \ if \ T-1 \leq t \leq T+P)  \}$. Similarly we have $(\gamma_{1},\Gamma_{1}), (\gamma_{2},\Gamma_{2}), (\eta_{1},\mathcal{H}), (a_{3},\mathcal{A}).$

\subsubsection{Main Algorithm}
Firstly, we apply Algorithm \ref{alg:nge} to train model (\ref{fir}) for obtaining $\hat{\beta}(t),\hat{\gamma}_1(t),\hat{\gamma}_2(t),$ $ \hat{\eta}_{1}(t)$ and $\hat{a}_{3}(t)$. Secondly, by using the following, we can estimate $\hat{Q}(t), \hat{I}(t) ,$ $\hat{A}(t),\hat{R}(t),\hat{D}(t)\hbox{ for }t \geq T-1$:

\begin{footnotesize}
\begin{equation} \label{pred SQIARD}
\begin{cases}
    \hat{S}(t+1) &= S(t) - \displaystyle \frac{\beta(t)S(t) (I(t) + \delta A(t))}{S(t) + I(t) + A(t)} + a_{3}(t) \mu_{2}Q(t)\\
    \hat{Q}(t+1) &= Q(t) + \displaystyle \frac{\beta(t)S(t) (I(t) + \delta A(t))}{S(t) + I(t) + A(t)}\\
    &- (a_{1}(t) \mu_{1}+a_{2}(t) \mu_{1}+a_{3}(t) \mu_{2}) Q(t)\\
    \hat{I}(t+1) &= I(t) + a_{1}(t) \mu_{1}Q(t) - \gamma_{1}(t)I(t) - \eta_{1}(t)I(t)\\
    \hat{A}(t+1) &= A(t) + a_{2}(t) \mu_{1}Q(t) - \gamma_{2}(t)A(t) - \eta_{2}(t)A(t)\\
    \hat{R}(t+1) &= R(t) + \gamma_{1}(t)I(t) + \gamma_{2}(t)A(t)\\
    \hat{D}(t+1) &= D(t) + \eta_{1}(t)I(t) + \eta_{2}(t)A(t).
\end{cases}
\end{equation}
\end{footnotesize}

\begin{algorithm}
\caption{Tracking Multiple SQIARD Models}
\label{alg:trackingSQIARD}
\begin{algorithmic}
\STATE{\textbf{Input:} Data: $\{Q(t), I(t), A(t),R(t),D(t), \ 0\leq t\leq T-1 \}$; Regularization Parameters: $m_{1}, m_{2}, m_{3}, m_{4}, m_{5}$; Orders of FIR: $J_1,J_2,J_3,J_4,J_5$; Criteria: P.}
\STATE{Calculate $\{ \beta(t),\gamma_{1}(t), \gamma_{2}(t),\eta_{1}(t), a_{3}(t),  \ 0 \leq t \leq T-2 \}$ by (\ref{SQIRDmodels}) and append to $ \mathcal{B}, \Gamma_{1}, \Gamma_{2}, \mathcal{H}, \mathcal{A}$, respectively.}
\STATE{Train models with (\ref{ridge}); $J_{i}, \ 1 \leq i \leq 5$; $m_{i}, \ 1 \leq i \leq 5$ and $ \mathcal{B}, \Gamma_{1}, \Gamma_{2}, \mathcal{H}, \mathcal{A}$, respectively by Algorithm \ref{alg:nge}}
\STATE{Estimate $\hat \beta(T-1),\hat \gamma_{1}(T-1),\hat \gamma_{2}(T-1),\hat \eta_{1}(T-1), \hat a_{3}(T-1)$ by (\ref{fir}), and append to $ \mathcal{B}, \Gamma_{1}, \Gamma_{2}, \mathcal{H}, \mathcal{A}$, respectively.}
\STATE{Estimate $\hat Q(T),\hat I(T),\hat A(T),\hat R(T),\hat D(T)$ by (\ref{pred SQIARD}).}
\WHILE{$T\leq t\leq T+P$}
\STATE{Train models with (\ref{ridge}); $J_{i}, \ 1 \leq i \leq 5$; $m_{i}, \ 1 \leq i \leq 5$ and $ \mathcal{B}, \Gamma_{1}, \Gamma_{2}, \mathcal{H}, \mathcal{A}$, respectively by Algorithm \ref{alg:nge}.}
\STATE{Estimate $\hat \beta(t),\hat \gamma_{1}(t),\hat \gamma_{2}(t),\hat \eta_{1}(t), \hat a_{3}(t)$ by (\ref{fir}), and append to  $ \mathcal{B}, \Gamma_{1}, \Gamma_{2}, \mathcal{H}, \mathcal{A}$, respectively.}
\STATE{Estimate $\hat Q(t+1),\hat I(t+1),\hat A(t+1),\hat R(t+1),\hat D(t+1)$ by (\ref{pred SQIARD}).}
\ENDWHILE
\RETURN Appended training data set: $ \mathcal{B}, \Gamma_{1}, \Gamma_{2}, \mathcal{H}, \mathcal{A}$; Predictions of $Q, I, A, R, D$: $\{\hat Q(t), \hat I(t), \hat A(t),\hat R(t),\hat D(t), \ T \leq t \leq T+P \}$.
\end{algorithmic}
\end{algorithm}

\newpage
\section{SIARD Model}
In the previous section, we established and discussed the SQIARD model. In order to implement the forecast for the most countries which don't provide the daily data for the quarantined, therefore we construct a new model in this section.
\subsection{The Derivation and Basic Reproduction Number}
In order to verify the epidemic effect of prediction  of the SIARD model, we will take the following two steps:
\begin{enumerate}
    \item Remove the parameter Q(t), and simplify the SQIARD infectious disease mathematical model under the other assumptions unchanged. Use the same training method to train the SIARD model and observe its effect of prediction .
    \item Use data from countries that have ``data on daily quarantine population" to compare the effect of prediction s of the two models on the epidemic.
\end{enumerate}

The variables are given as follows: [$S$: susceptible population; $I$: infective population; $A$: asymptomatic infective population; $R$: recovered population; $D$: deaths].
\begin{figure}[htp]
    \centering
    \includegraphics[width=8 cm]{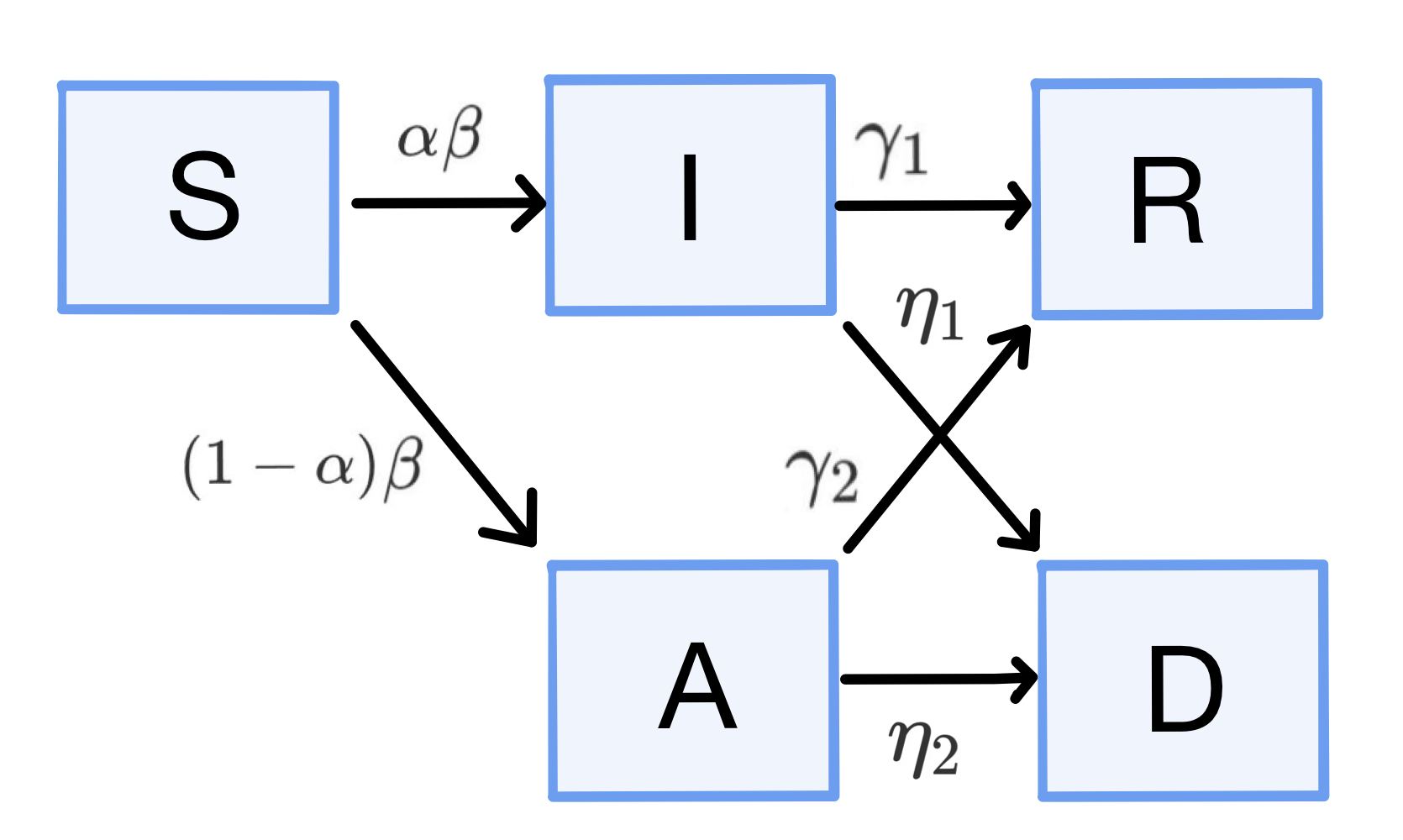}\\
    \caption{The flow chart of SIARD}
\end{figure}
The model parameters are given as follows: [$\beta$: the progression rate of susceptible class to infective classes;
$\delta$: the reduction in infectiousness of asymptomatic infectives, where $0<\delta<1$; $\alpha$: the fraction of susceptible from susceptible to $I$ or $A$, where $0<\alpha<1$; $\gamma_1$ and $\gamma_2$: the recovered rates of infective classes $I$ and $A$; $\eta_1$ and $\eta_2$: are the disease death rates of infective classes $I$ and $A$].

\begin{equation} \label{eq:SIARD}
    \begin{aligned}
    S' & =-\displaystyle\frac{\beta S(I+\delta A)}{S+I+A}\\
    I' & =\alpha\displaystyle\frac{\beta S(I+\delta 
    A)}{S+I+A} -\gamma_1 I -\eta_1 I\\
    A' & =(1-\alpha)\displaystyle\frac{\beta S(I+\delta A)}{S+I+A} -\gamma_2 A -\eta_2 A\\
    R' & = \gamma_1 I+\gamma_2 A \\
    D' & = \eta_1 I + \eta_2 A\\
    \end{aligned}
\end{equation}

$S(0)=S_0>0$ , $I(0)=I_0>0$ , $A(0)=A_0>0$ , $R(0)=0$ and $D(0)=0$.\\
We note that $S(t)+I(t)+A(t)+R(t)+D(t)=N=S_0+I_0+A_0$ where $N$ is the total population.
Note that $R_0$ is simply the basic reproduction number of this system.
To further examine the stability condition of such a system , we let 
$$R_0=\alpha\displaystyle\frac{\beta}{\gamma_1+\eta_1}+(1-\alpha)\displaystyle\frac{\beta\delta}{\gamma_2+\eta_2}.$$

Due to the COVID-19 data is uploaded in days, we revise the differential equation into discrete time difference equation.
\begin{center}
\begin{equation}
    \begin{array}{l}
    S(t+1)-S(t)=-\displaystyle\frac{\beta S(I+\delta A)}{S+I+A}\\
    \ \\
    I(t+1)-I(t)=\alpha\displaystyle\frac{\beta S(I+\delta 
    A)}{S+I+A} -\gamma_1 I -\eta_1 I\\
    \ \\
    A(t+1)-A(t)=(1-\alpha)\displaystyle\frac{\beta S(I+\delta A)}{S+I+A} -\gamma_2 A -\eta_2 A\\
    \ \\
    R(t+1)-R(t) = \gamma_1 I+\gamma_2 A \\
    \ \\
    D(t+1)-D(t) = \eta_1 I + \eta_2 A\\
    \end{array}
\end{equation}
\end{center}
When the disease first spreads, the number of people infected is much smaller than the total population, the number of suspected infections is approximated to the total of population. Then above equations can simplified as follows :
\begin{equation}
    \begin{aligned}
    I(t+1)-I(t) & =\alpha\beta (I+\delta 
    A)-\gamma_1 I -\eta_1 I\\
    A(t+1)-A(t) & =(1-\alpha)\beta (I+\delta A)-\gamma_2 A -\eta_2 A\\
    \end{aligned}
\end{equation}

\subsection{Tracking Time-Depend SIARD Model Algorithms}
For the SIARD, $\beta(t), \gamma_{1}(t), \gamma_{2}(t), \eta_{1}(t)$ can be evolved as (\ref{SIARDmodels}) from the discrete form of the SIARD differential equation.
\begin{scriptsize}
\begin{equation} \label{SIARDmodels}
\begin{aligned}
    \beta(t) & = \frac{S(t) - S(t+1)}{S(t)(I(t) + \delta A(t))} (S(t) + I(t) + A(t))\\
    \gamma_{1}(t) & = (1+\frac{D(t+1)-D(t)}{I(t)}) + \alpha \frac{S(t) - S(t+1)}{I(t)} - \frac{I(t+1)}{I(t)}\\
    \gamma_{2}(t) & = 1 + (1 - \alpha) \frac{S(t) - S(t+1)}{A(t)} - \frac{A(t+1)}{A(t)}\\
    \eta_{1}(t) & = \frac{D(t+1)-D(t)}{I(t)}
\end{aligned}
\end{equation}
\end{scriptsize}

Similarly, in order to estimating $\hat{I}(t),\hat{A}(t),\hat{R}(t),\hat{D}(t)$ with SIARD for $t > T$, we use the Algorithm \ref{alg:nge} to train models of (\ref{fir}) without $\hat{a}_{3}$ and obtain $\hat{\beta} (t)$, $\hat{\gamma_{1}} (t)$, $\hat{\gamma_{2}} (t)$, $\hat{\eta_{1}} (t)$. Then, we use it to compute $\hat{I}(t) ,\hat{A}(t),\hat{R}(t),\hat{D}(t)$ as (\ref{pred SIARD}) and also append the prediction $\hat{\beta} (t),\hat{\gamma_{1}} (t),\hat{\gamma_{2}} (t),\hat{\eta_{1}} (t), t \geq T-1$ to $\mathcal{B}$,$\Gamma_{1}$,$\Gamma_{2}$,$\mathcal{H}$ respectively.
\begin{small}
\begin{equation} \label{pred SIARD}
\begin{cases}
    \hat{S}(t+1) &= S(t) - \displaystyle \frac{\beta(t)S(t) (I(t) + \delta A(t))}{S(t) + I(t) + A(t)}\\
    \hat{I}(t+1) &= I(t) + \alpha \displaystyle \frac{\beta(t)S(t) (I(t) + \delta A(t))}{S(t) + I(t) + A(t)} - \gamma_{1}(t)I(t) \\
    &- \eta_{1}(t)I(t)\\
    \hat{A}(t+1) &= A(t) + (1-\alpha) \displaystyle \frac{\beta(t)S(t) (I(t) + \delta A(t))}{S(t) + I(t) + A(t)} - \gamma_{2}(t)A(t)\\
    &- \eta_{2}(t)A(t)\\
    \hat{R}(t+1) &= R(t) + \gamma_{1}(t)I(t) + \gamma_{2}(t)A(t)\\
    \hat{D}(t+1) &= D(t) + \eta_{1}(t)I(t) + \eta_{2}(t)A(t),
\end{cases}
\end{equation}
\end{small}
where $t \geq T-1$. Also, before implementing the Tracking Algorithm, we have to conduct the orders by Algorithm \ref{alg:ordersearcher} first, then obtaining Tracking SIARD Algorithm from revised Algorithm \ref{alg:trackingSQIARD} by removing the $Q(t)$, $\hat{Q}(t)$, $a_{3}(t)$, $\hat{a}_{3}(t)$. Section 4 is our result of the forecast.

\newpage
\section{Implement and Numerical Analysis}
In this section, we apply SIARD model to US, Brazil, South Korea, India, Russia, Italy, and SQIARD model to US with data sets \cite{data1},\cite{data2} and \cite{data3}, then showing the result in each country. The following Table 1 is the parameters for the forecast in SQIARD and SIARD.

\begin{table}[htbp]
{\scriptsize
  \caption{Parameters for SQIARD and SIARD } \label{tab:par}
\begin{center}
  \begin{tabular}{||c||c c c c c c c||}
    \hline
    \hline
    Parameters & $ S_0$&$\alpha$&$\beta_o$&$\gamma_{1o}$&$\gamma_{2o}$&$\eta_o$&$a_{3o}$\\
    \hline
    US (SQ)&328200000&0.4&8&3&3&15&8\\
    \hline
    US (SI)&328200000&0.4&3&3&3&15& None\\
    \hline
    Brazil (SI)&209500000&0.4&12&19&19&5& None\\
    \hline
    South Korea (SI)&51640000&0.4&5&3&3&14& None\\
    \hline
    India (SI)&1353000000&0.4&3&16&16&3& None\\
    \hline
    Russia (SI)&145500000&0.6&17&7&7&3& None\\
    \hline
    Italy (SI)&60360000&0.55&17&5&5&12& None\\
    \hline
    \hline
  \end{tabular}
\end{center}
}
\end{table}

\subsection{Forecast of SQIARD in US}
For the SQIARD model, we consider the case $\mu_1 =1$ and\ $\mu_2=0.14$ where $\mu_1$ and\ $\mu_2$ are the rate from quarantine to I and A. In Figure 4, first, we use 100 data to train model and 20 validation data to find the best fitting orders of respective FIR model, then, we obtain the reference effective intervals with range 9 days in 5 \% relative error, 18 days in 10 \% relative error and 20 days in 20 \% relative error from SQIARD model.

\begin{figure}[htp]
    \centering
    \includegraphics[width=10 cm]{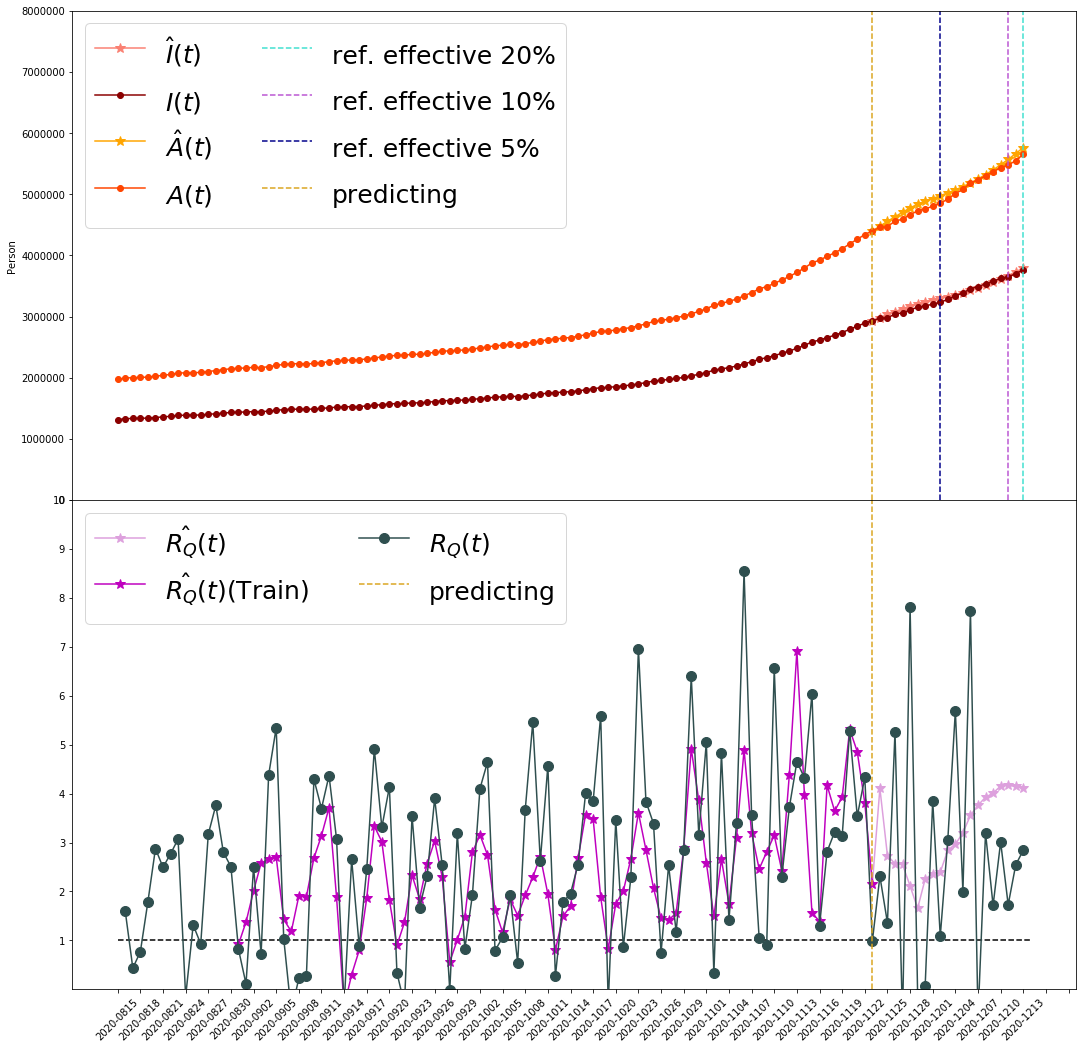}\\
    \caption{Find the best fit order of $a_{3}$, $\beta$, $\gamma_{1}$,$\gamma_{2}$, $\eta_{1}$,$\eta_{2}$ depending on the validation set with size 20.}
\end{figure}

\newpage
\begin{figure}[htp]
    \centering
    \includegraphics[width=10 cm]{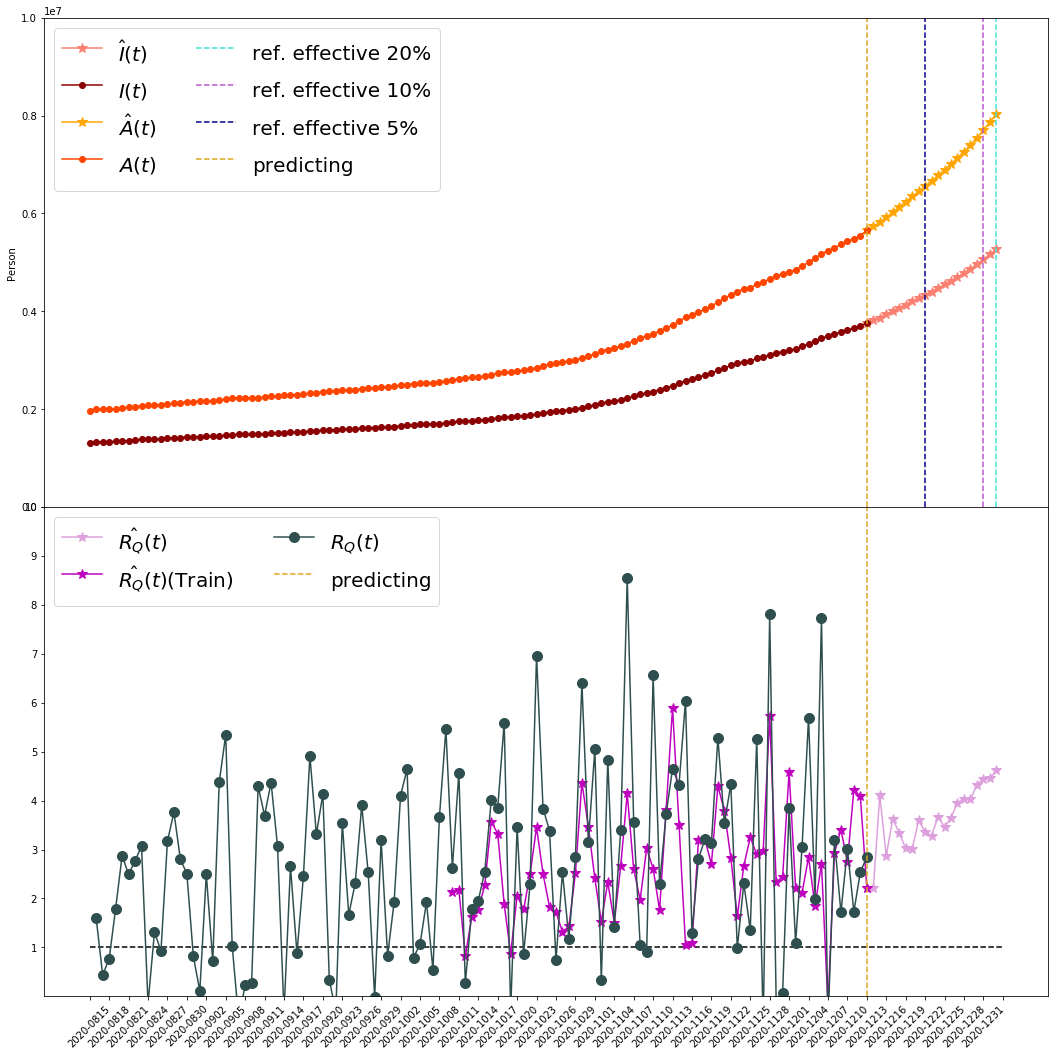}\\
    \caption{Use the found orders from Fig. 4 for each model to conduct the forecast for 20 days in the future.}
\end{figure}

In Figure 5, we use the trained model to predict 20 days and compare to the validation data, and applying three kinds of relative error, 5 \%, 10 \% and 20 \% to compute the amount of days which in relative error respectively. Then, we take these days as our reference effective interval. For example, in Figure 4, we obtain the reference effective interval with range 9 days in 5 \% relative error, 18 days in 10 \% relative error and 20 days in 20 \% relative error. And, we expand our training data to 120 data and predict 20 data in the future.

\newpage
\subsection{Forecast of SIARD}
On the other hand, for the SIARD model, we obtain the reference effective interval with range 11 days in 5 \% relative error and 20 days in 10 \% relative error from SIARD model.
We also can see the trend of I, A are getting slow down, so the value of$\ R_0$ is getting smaller.
\begin{figure}[tbhp]
\centering
\subfloat[US SIARD]{\label{fig:SIARD_US}\includegraphics[width=5.5cm]{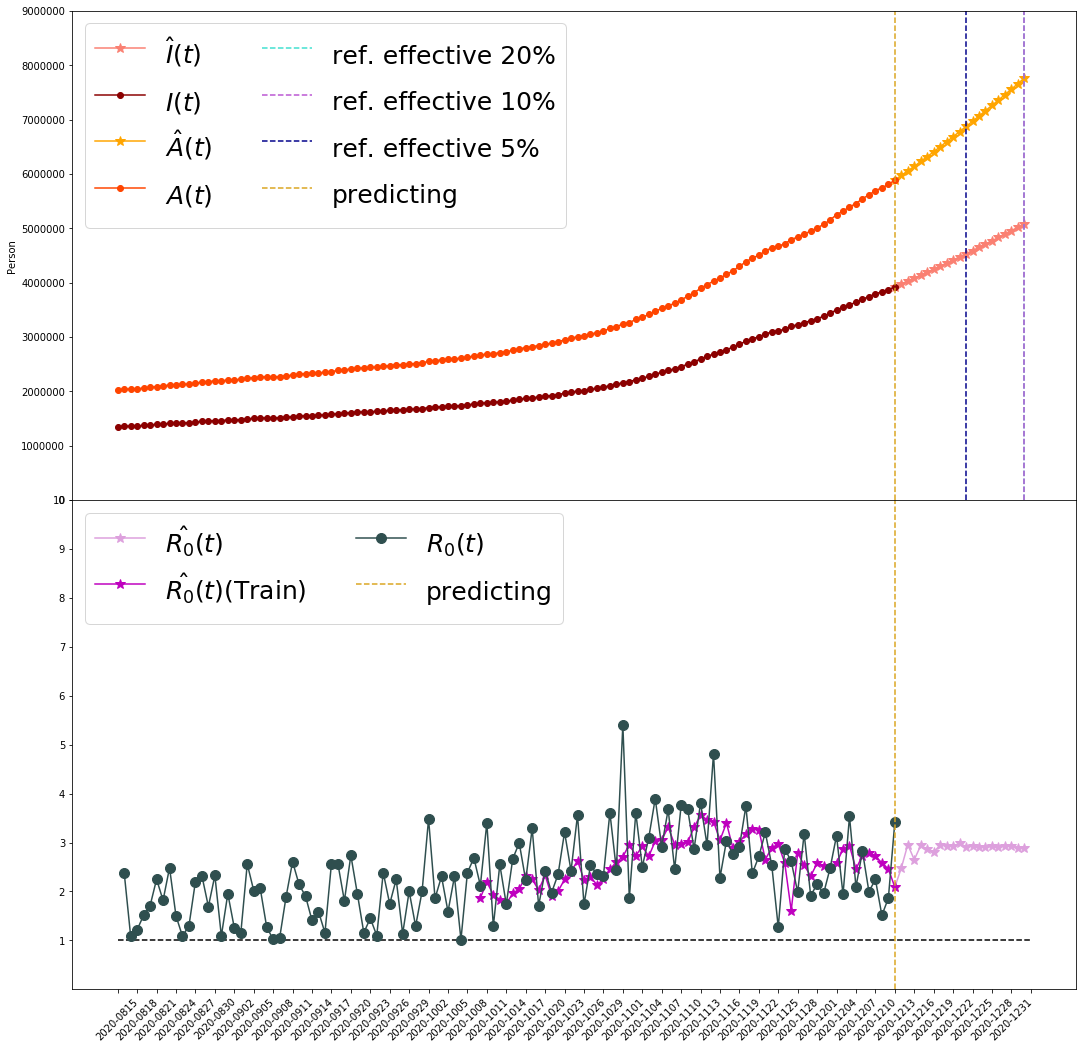}}
\subfloat[Brazil SIARD]{\label{fig:SIARD_Brazil}\includegraphics[width=5.5cm]{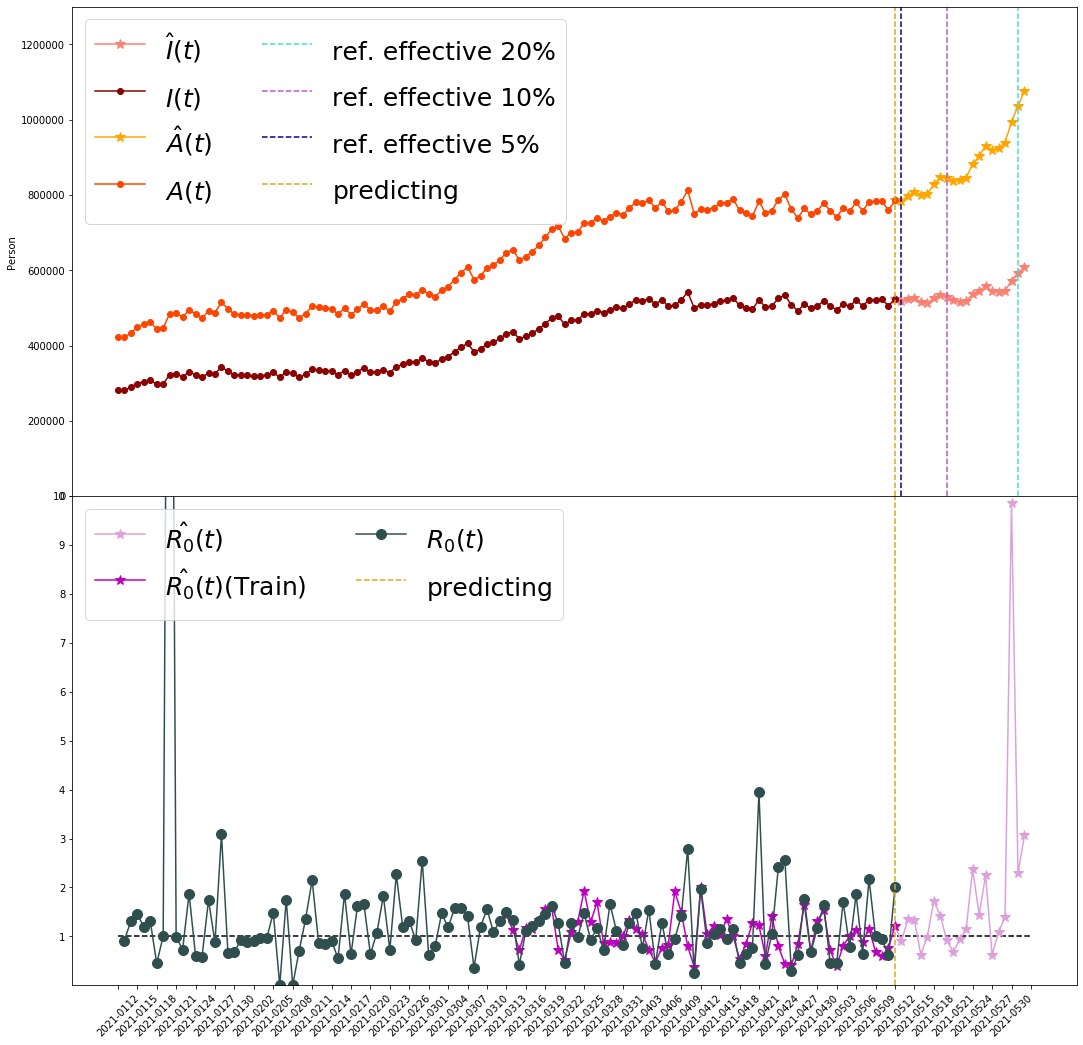}}
\caption{SIARD US and Brazil.}
\end{figure}

\begin{figure}[tbhp]
\centering
\subfloat[South Korea SIARD]{\label{fig:SIARD_Korea}\includegraphics[width=5.5cm]{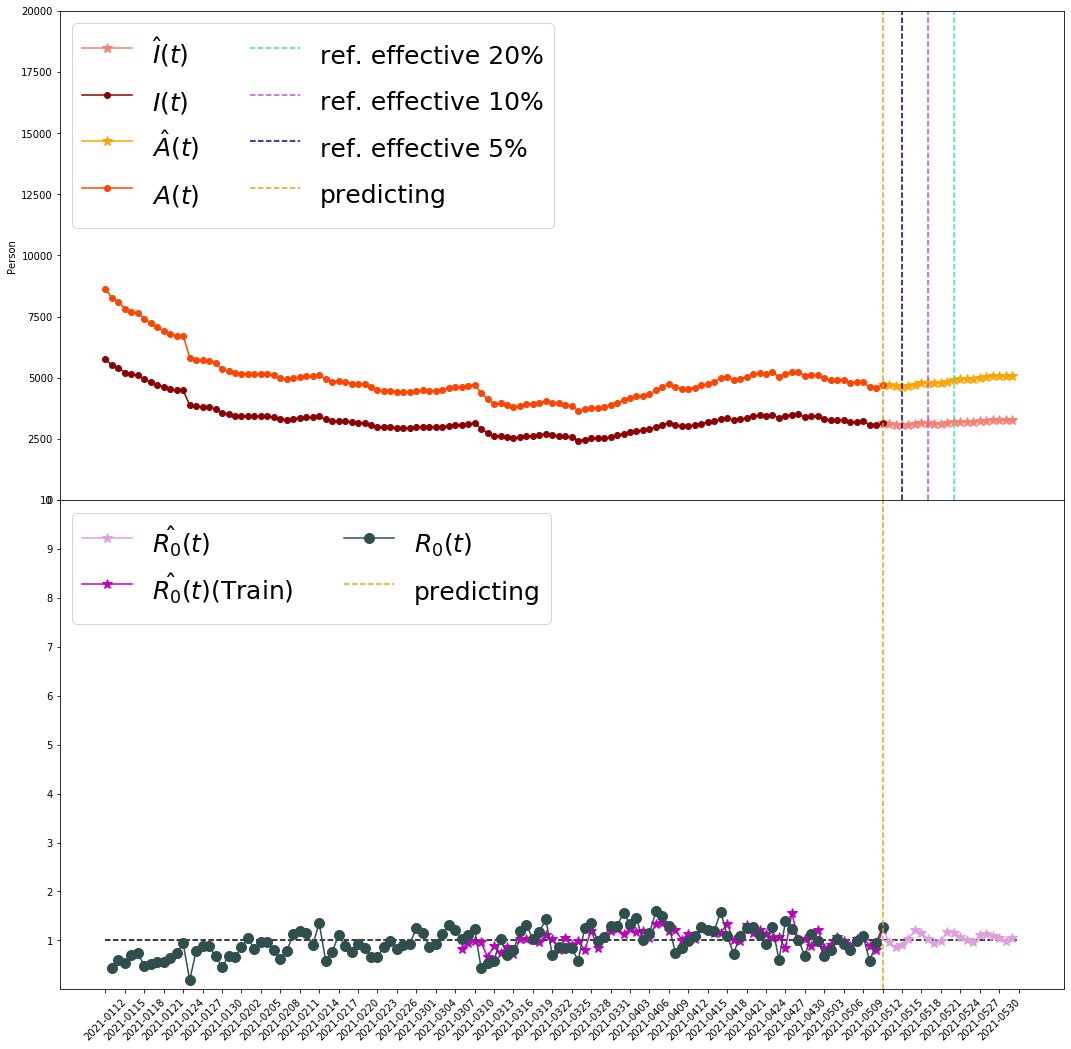}}
\subfloat[India SIARD]{\label{fig:SIARD_India}\includegraphics[width=5.5cm]{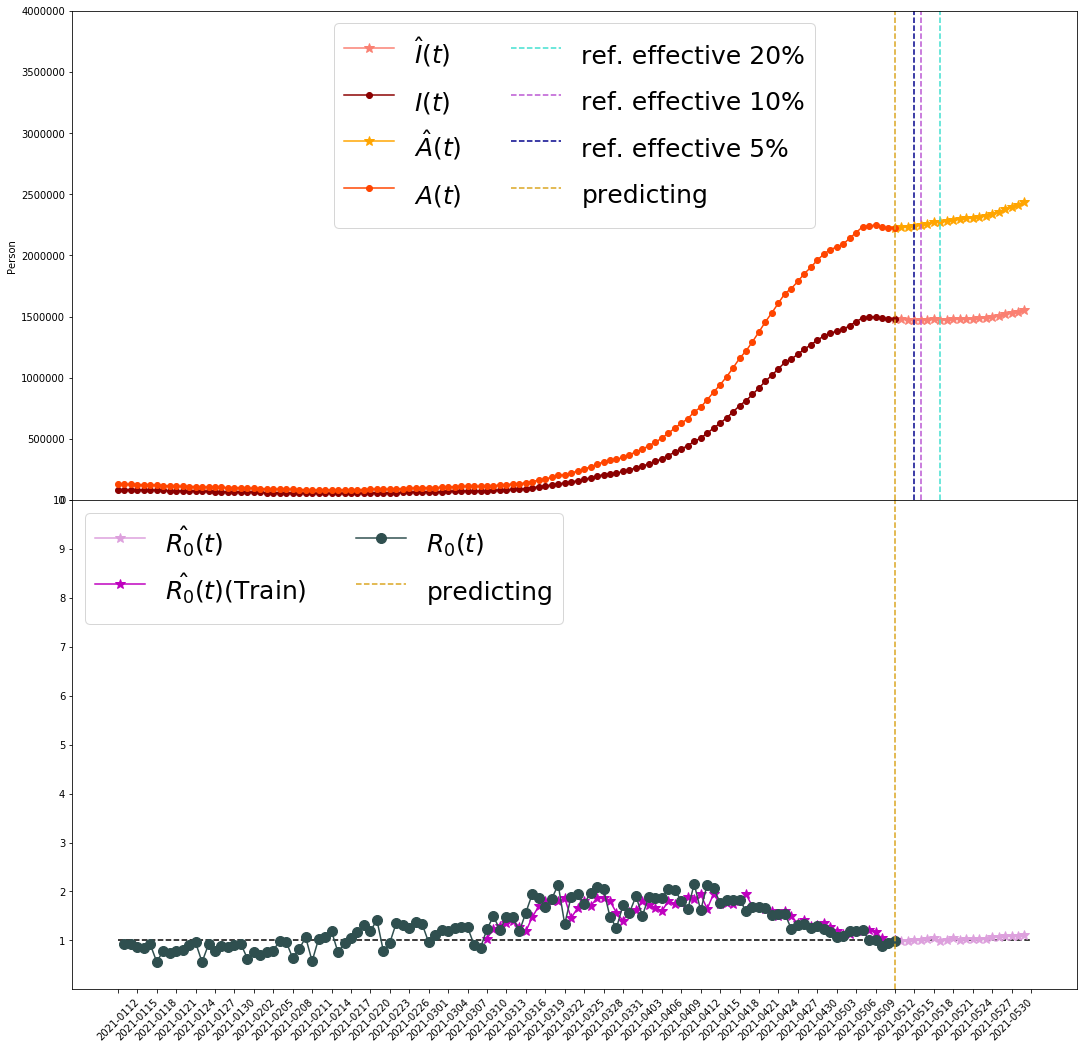}}
\caption{SIARD South Korea and India.}
\end{figure}

\begin{figure}[tbhp]
\centering
\subfloat[Russia SIARD]{\label{fig:SIARD_Russia}\includegraphics[width=5.5cm]{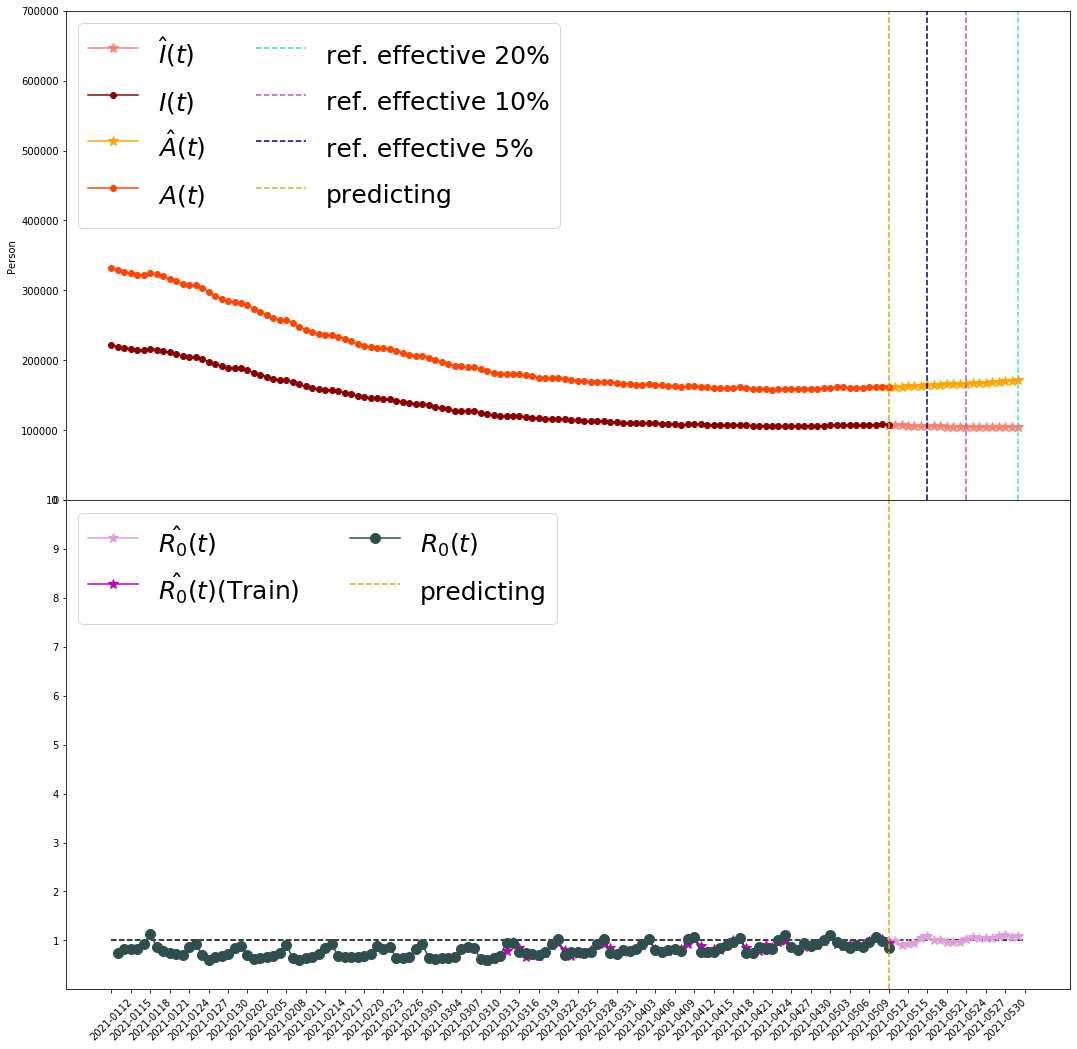}}
\subfloat[Italy SIARD]{\label{fig:SIARD_Italy}\includegraphics[width=5.5cm]{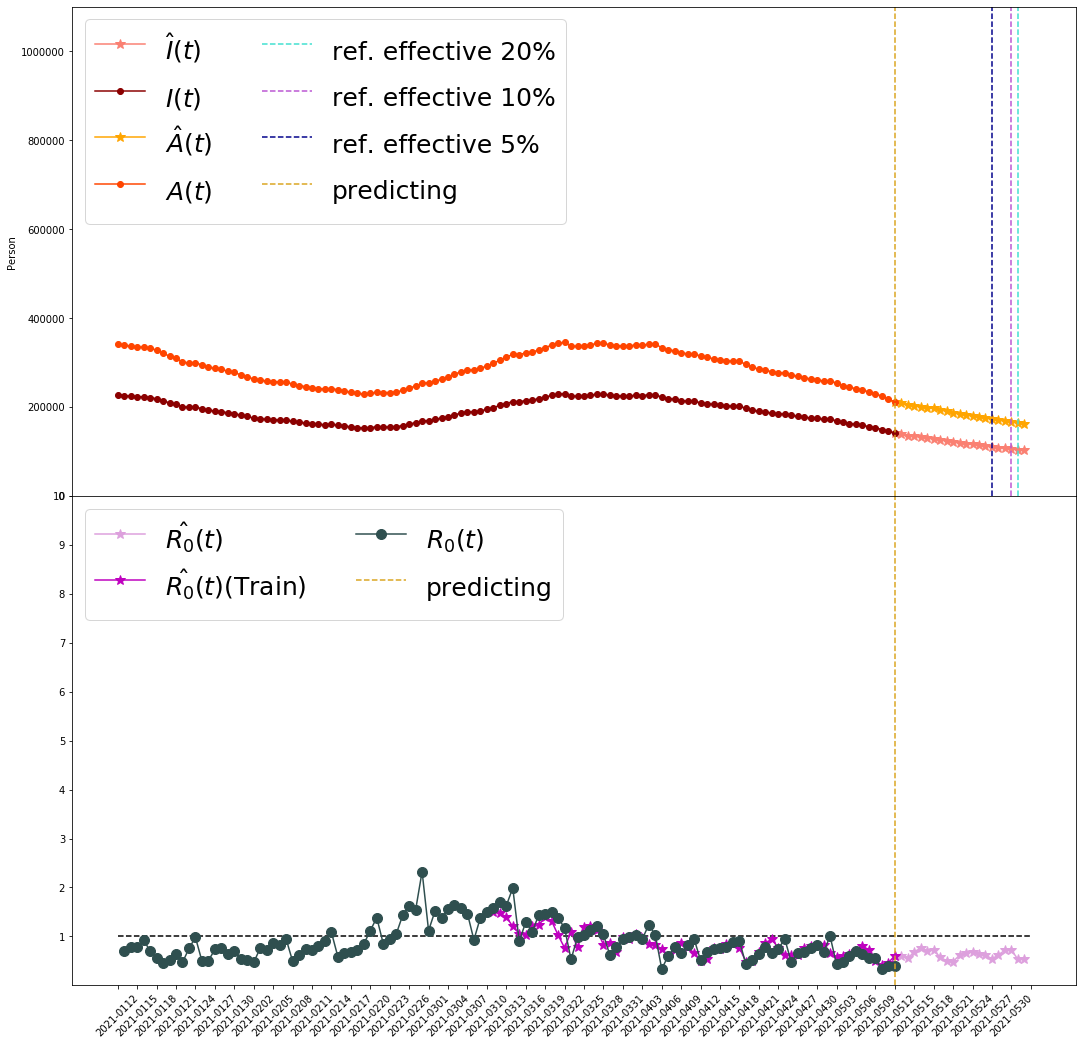}}
\caption{SIARD Russia and Italy.}
\end{figure}

In Brazil (FIG. \ref{fig:SIARD_Brazil}), we obtain the reference effective interval: 1 day in 5\% relative error, 10 days in 10 \% relative error and 14 days in 20 \% relative error. 

In South Korea (FIG. \ref{fig:SIARD_Korea}), we obtain the reference effective interval: 1 day in 5\% relative error, 2 day in 10 \% relative error and 7 days in 20 \% relative error. We can see I, A in middle part are getting higher, same as$\ R_0$. Conversely, I, A decrease in last part, then$\ R_0$ is getting lower and most of$\ R_0$ are lower than 1, so the epidemic in Korea may be controlled. 

In India (FIG. \ref{fig:SIARD_India}), we obtain the reference effective interval: 5 days in 5\% relative error, 20 days in 10 \% relative error.

\newpage
In Russia (FIG. \ref{fig:SIARD_Russia}), we obtain the reference effective interval: 15 days in 5\% relative error,
20 days in 10 \% relative error, it has lots of days in 5 \% relative error in six countries. In other words, the trained model of Russia has caught the trend of data.

In Italy (FIG. \ref{fig:SIARD_Italy}), we obtain the reference effective interval: 4 days in 5\% relative error, 20 days in 10 \% relative error. The trend of I, A are decrease, Covid-19 may be controlled in Italy.

\subsection{Forecast Result from Nov.23 to Dec.13, 2020 for US, and Apr.22 to May.10, 2021 for Others five Countries}
Since the data about quarantine of US interrupted on Dec.13, 2020, we compare with real data from Nov.23 to Dec.13, 2020 in US. The followings are the results about the forecast for SQIARD and SIARD, see Figure 9.

\begin{figure}[tbhp]
\centering
\subfloat[US SQIARD]{\includegraphics[width=5.5cm]{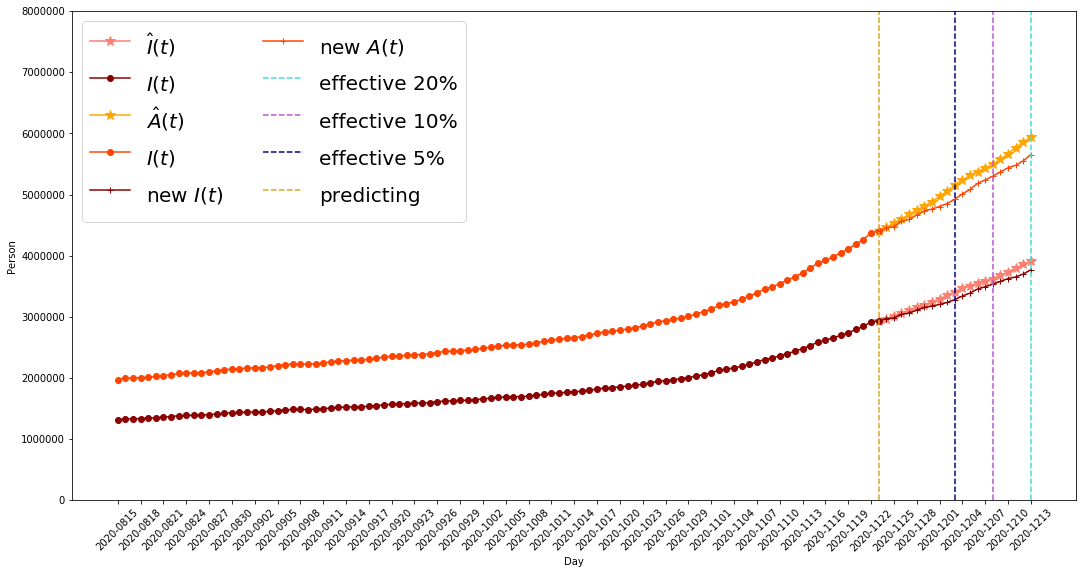}}
\subfloat[US SIARD]{\includegraphics[width=5.5cm]{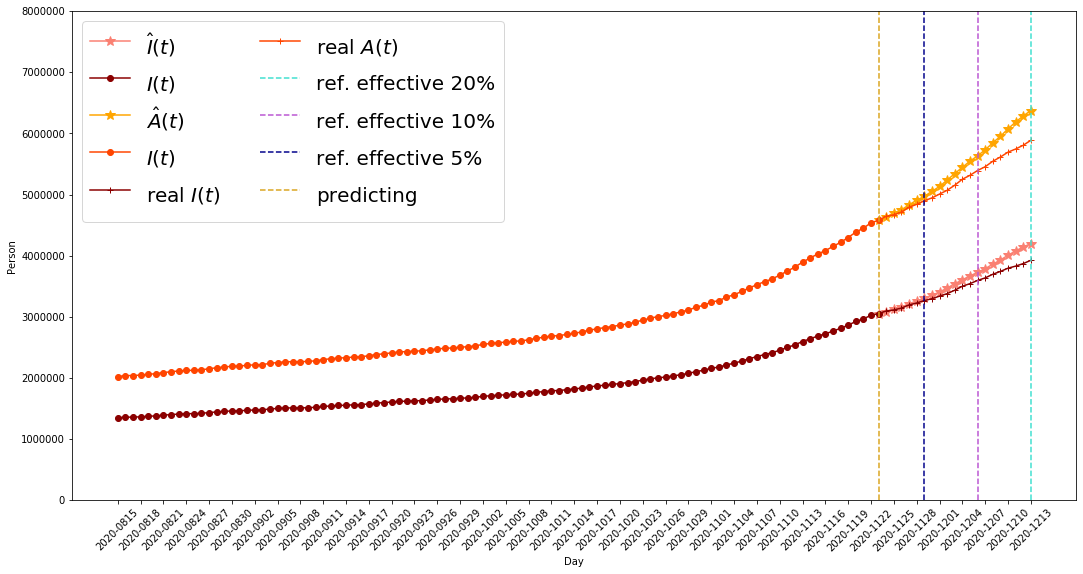}}
\caption{Forecast of SQIARD and SIARD in US from Nov.23 to Dec.13, 2020}
\end{figure}

In US, we have the data of quarantine, so we can apply SQIARD model to predict data in future. Since we add extra parameters into model, especially the speed rate from quarantine to infected or asymptomatic infected, the data can be predicted more precisely than SIARD model. In Fig. 9, we found that, the data predicted by SQIARD model are more closed to the real data.

Simultaneously, we use SIARD model to conduct forecast, the reference effective interval has 6 days in 5 \% relative error(see Fig. 9(b)). When we use SQIARD model to predict, the reference effective interval has 9 days in 5 \% relative error(see Fig. 9(a)), so the effect of SQIARD model is better than SIARD model during this period.

\newpage
The following Figure 10 are the results about the forecast for SIARD from  Apr.22 to May.10. 2021 in Brazil, South Korea, India, Russia and Italy.
\begin{figure}[tbhp]
\centering
\subfloat[Brazil SIARD]{\includegraphics[width=5.5cm]{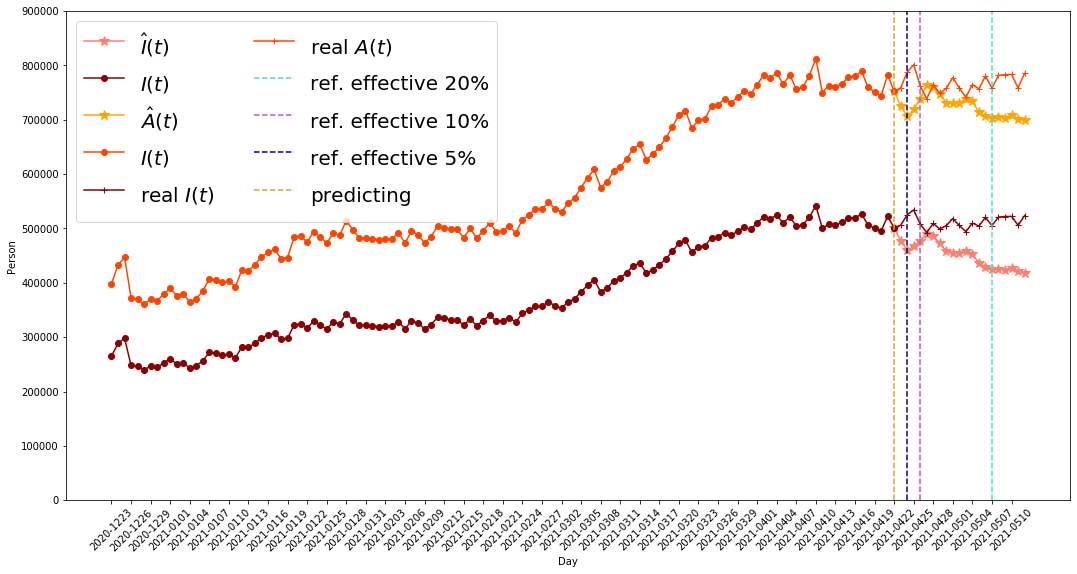}}
\subfloat[South Korea SIARD]{\includegraphics[width=5.5cm]{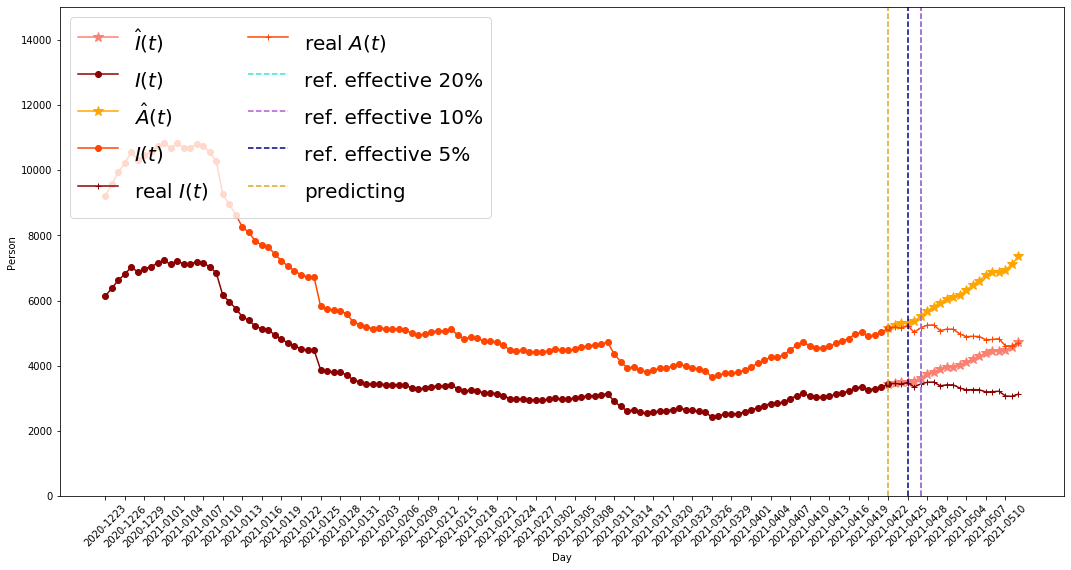}}
\ \\
\subfloat[India SIARD]{\includegraphics[width=5.5cm]{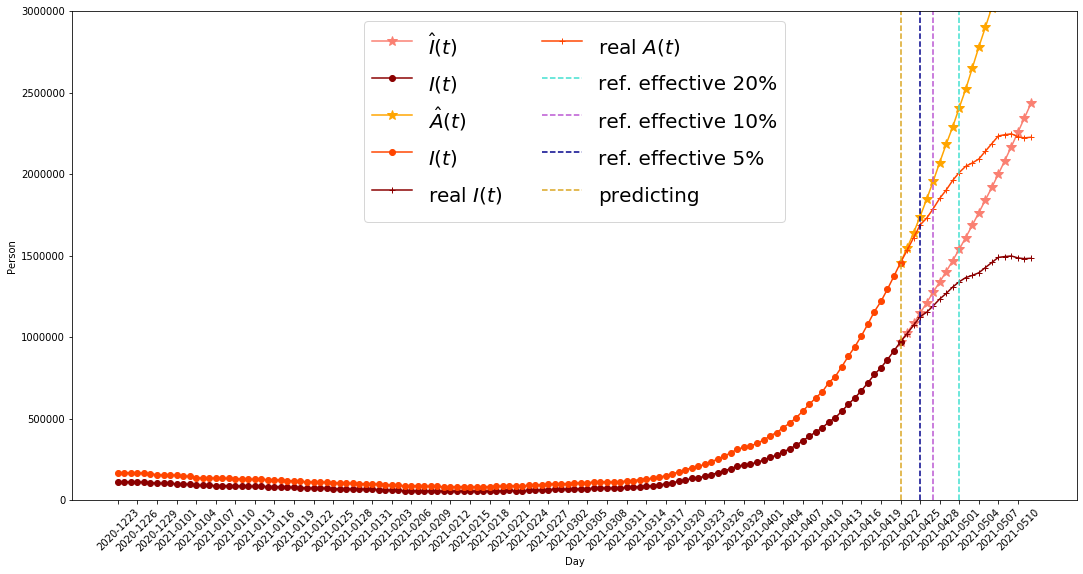}}
\subfloat[Russia SIARD]{\includegraphics[width=5.5cm]{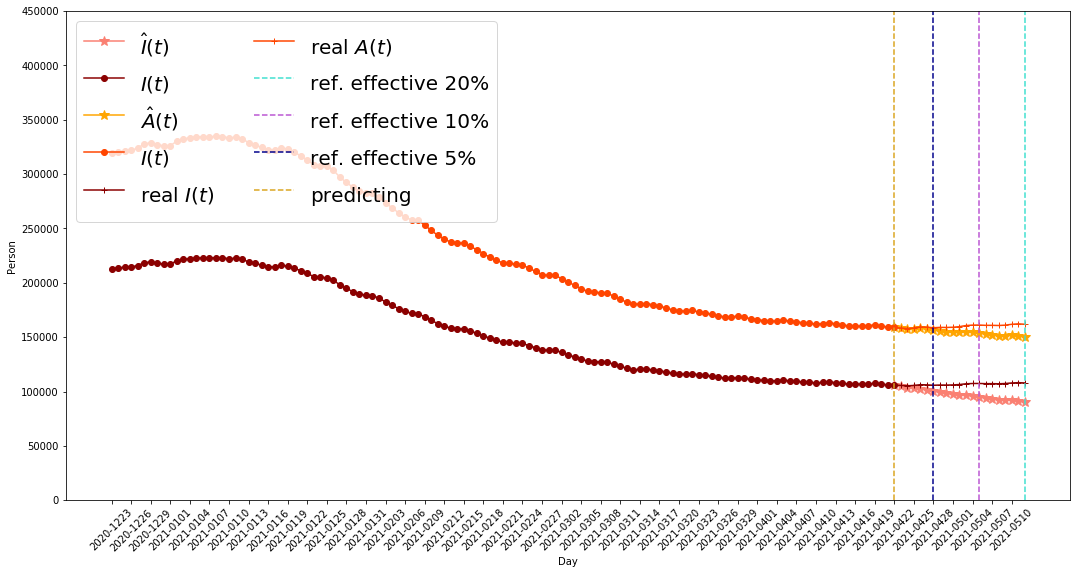}}
\ \\
\subfloat[Italy SIARD]{\includegraphics[width=5.5cm]{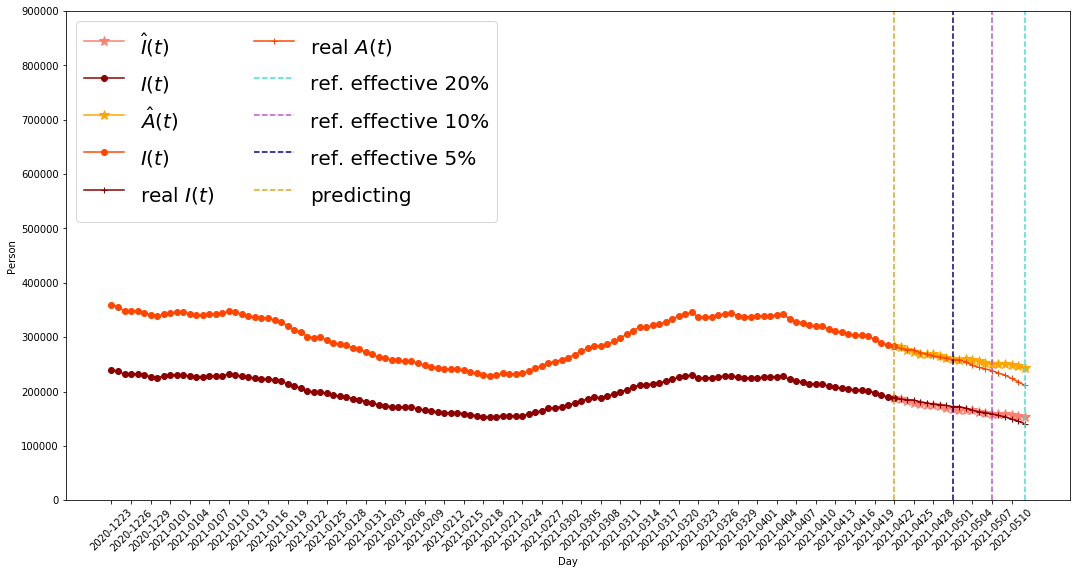}}
\caption{Forecast of SIARD in Brazil, South Korea, India, Russia and Italy from  Apr.22 to May.10. 2021}
\label{fig:Jan.12 to Jan.31.}
\end{figure}


\subsection{Comparison and Analysis of Forecasting Effects}

Our model finds the proportion of people with symptoms $\alpha$ in six countries. For example, in Table 1, the proportion of symptomatic infections in Italy is about 55 \%, which means that the proportion of asymptomatic infections in the country is about 45 \%. This is in line with the 43.2 \% of asymptomatic infections obtained by the authors of \cite{intro2} at V`o, Italy. It proves that our model does have the ability to judge the proportion of asymptomatic infections.

By above figures in section 4, it is obvious that the trend of symptomatic infections and asymptomatic infections are in relation with$\ R_0$. When$\ R_0$ increases, I and A also increase, same as decrease, so the result of our prediction is accord with the definition of$\ R_0$. Hence, from our data prediction, it also show$\ R_0$ can be viewed as an important target of the break or not of the Covid-19 epidemic.





\end{document}